\documentstyle[aps,twocolumn]{revtex}

\begin{document}
\title{Anomalous Parallel Field Negative Magnetoresistance in Ultrathin Films near
the Superconductor-Insulator Transition}
\author{Kevin A. Parendo, L. M. Hernandez, A. Bhattacharya, and A. M. Goldman}
\address{School of Physics and Astronomy, University of Minnesota, 116 Church St. SE,%
\\
Minneapolis, MN 55455, USA}
\date{
\today%
}
\maketitle

\begin{abstract}
A parallel field negative magnetoresistance has been found in
quench-condensed ultrathin films of amorphous bismuth in the immediate
vicinity of the thickness-tuned superconductor-insulator transition. The
effect appears to be a signature of quantum fluctuations of the order
parameter associated with the quantum critical point.
\end{abstract}

\pacs{PACS numbers: ()}

Quantum phase transitions (QPTs) are brought about by the variation of an
external parameter of the Hamiltonian of a system, which changes the ground
state \cite{Girvin}. The superconductor-insulator (SI) transition in two
dimensions (2D), tuned by disorder or magnetic field, is believed to be a
quantum phase transition. The understanding of the SI transition as a QPT
has been inferred from the successful analysis of transport data using
finite size scaling. However recent data on the field- and disorder-tuned SI
transitions suggest the existence of a finite intermediate regime of
metallic behavior not anticipated by the theory \cite{Mason}. Subsequent
explanations of this regime have included a metallic Bose glass or a Bose
metal \cite{Phillips}, metallicity produced by the influence of dissipation 
\cite{Mason}, and effects resulting from the influence of fermionic
excitations not included in boson models \cite{Trivedi}. In some instances
the metallic regime may be attributable to the electrons not cooling.
Because of these complications, it would be useful if there were an explicit
signature of quantum fluctuations that could serve as an indicator of the SI
transition. A recent calculation\cite{Lopatin} appears to offer this
possibility. Employing a perturbative approach, a negative correction to the
parallel field magnetoresistance (MR) attributable to quantum fluctuations
has been found near the parallel-field SI transition of films (and wires).
The total negative MR results from the ``Aslamazov-Larkin'' correction being
overwhelmed by negative contributions from the ``density of states'' and
``Maki-Thompson'' terms. In this letter we report an anomalous, {\it parallel%
}-field negative MR whose occurrence is correlated with the thickness-tuned
SI transition of ultrathin homogeneous films. This effect may derive from
corrections to the conductivity associated with quantum fluctuations even
though the effect is found near the condition of critical disorder rather
than critical parallel magnetic field.

Resistance measurements were made using a bottom loading Kelvinox 400
dilution refrigerator, employing four-probe techniques. Electrical leads
were filtered at room temperature using $\pi $-section filters with a cutoff
frequency of about 10 Hz. Power dissipation in the measurement process was
kept below 1 pW. The substrate was mounted on a sample holder that could be
transferred between the mixing chamber of the refrigerator and an attached
ultra-high vacuum growth chamber using a liquid-helium-cooled transfer stick 
\cite{Hernand}. In these experiments the plane of the substrate, mounted on
a rotatable sample holder, was restricted to be close to the nominally
parallel orientation to accommodate additional heat sinking needed to
facilitate cooling below 0.1K.

Films were grown on substrates held at liquid helium temperatures while
mounted on the sample stick with the growth chamber at a pressure of $%
10^{-10}$ Torr. The substrates were epi-polished single-crystal SrTiO$_{3}$%
(100) wafers, pre-coated ({\it in situ}) with a $6$\AA\ thick film of {\it a-%
}Ge. To prevent annealing, substrate temperatures were held below $12K$
during growth, and below $18K$ during other processing and handling. Film
thicknesses were increased in increments as small as $0.04$ \AA ,\ as
measured using a calibrated quartz crystal monitor. The latter was
calibrated {\it ex situ} using a profilometer. Films processed in this
fashion are believed to be homogeneously disordered and not granular \cite
{Strongin}. Critical features of the present experiments were the
possibilities of changing the thickness of a film in tiny increments and of
growing films homogeneous in thickness to one part in 10$^{4}\cite{Hernand}.$
The phenomena reported here occurred over a nominal thickness range of order
0.8\AA\ out of approximately 9.0\AA , and would not have been seen without
such stringent control.

The evolution of $R(T)$ of eleven films with thicknesses ranging from 8.5
\AA\ to 9.3 \AA\ is shown in Fig 1. Thinner and thicker films, grown in
other runs (not shown) were insulating and glass-like in their responses, or
fully superconducting, respectively. It should be noted that there are
metallic regimes at low temperatures in this data, for both insulator- and
superconductor-like films. From this set of experiments alone one cannot
demonstrate that these regimes are not a consequence of failure to cool the
electrons. However the existence of an intermediate metallic regime \cite
{Mason,Phillips,Chervenak,Doniach,Christiansen} separating superconducting
and insulating films is not the issue in the present work. Apart from this
possible metallic behavior at the lowest temperatures, the films sort into
two categories, insulator-like $(dR/dT<0)$ and superconductor like $%
(dR/dT>0) $.

Films with thicknesses less than $8.99\AA $ are in the insulating state and
have positive $R(B)$ at all temperatures. The negative MR first appears in
the 8.99\AA\ thick film and was studied at temperatures between 0.05 and
0.3K. An example of the temperature dependence of the MR for the 9.05\AA\
film, which is representative, is exhibited in Fig. 2. In the lowest fields, 
$dR/dB>0$. With increasing field, a maximum is reached. At all but the
highest temperatures, a regime in which $dR/dB<0$ is then entered. With
further increase in field, there is a minimum in $R(B)$, followed by a
regime in which the resistance is a linear function of field. This linear
behavior at high fields is found at all temperatures from 0.05K to 1K and in
fields from 2T to 12T.

The magnetic fields in these measurements were only nominally parallel to
the substrate plane. The misalignment was estimated to be at most the order
of 1 to 2$^{0}.$ This would lead to a perpendicular field component of about
1/30th that of the applied field. At low applied fields, the resultant
perpendicular component is insignificant. However, as the magnetic field is
increased, this will eventually no longer be true. At high fields we find a
linear dependence of $R(B),$ an expected effect if there were flux flow
resistance due to a perpendicular field component\cite{Markovic1}. With the
above estimate of the misalignment this linear regime would appear to start
at perpendicular field components of the order of 600 Oe.

There are systematic aspects of the low field data exhibiting positive MR,
which lead us to attribute it to a spin polarization of the carriers
transported by hopping in the {\it a-}Ge substrate. The effect is most
pronounced in the thinnest films, where contributions to the conductivity
from the substrate would be proportionally more important than in thicker
ones. The peak disappears entirely for films whose thickness exceeds 9.09\AA
. This would be expected when transport through the film became dominant. A
theory of positive MR in the hopping regime has been given by Matveev and
collaborators \cite{Matveev}. It is based on the idea that in zero magnetic
field a significant number of hopping sites can be doubly occupied with the
electrons forming a spin singlet. In a magnetic field strong enough to
polarize the carriers, transitions involving such sites are forbidden as
electron pairs cannot form singlets. This leads to a positive MR that
saturates when the spins were fully polarized. This picture has been
verified experimentally in semiconducting films\cite{Mertes}. In the films
of the present work, the conductance channel exhibiting low-field positive
MR competes with that exhibiting negative MR, which is a parallel channel.
When the positive MR saturates, the negative MR dominates, resulting in a
relatively sharp peak at the lowest temperatures for the thinnest films. The
peak field should occur when the condition $\mu _{B}B\sim k_{B}T$ is
satisfied. In Fig. 3 we plot the field at the MR peak vs. T for films of
three different thicknesses. The line on the figure corresponds to $\mu
_{B}B=k_{B}T.$

It is necessary to understand the systematics of the negative MR effect in
order to justify relating it to quantum fluctuations associated with a
quantum critical point. In Fig. 4 we plot the fractional change in
resistance from the peak to the trough of the negative magnetoresistance as
a function of film thickness at 0.050K, 0.200K and 0.3K. Measurements at
other temperatures have been suppressed for clarity. The negative MR is not
found in any of the films at 0.3K, and is strongest at the lowest
temperature, 0.050K, for the 9.19\AA\ thick film. It is first found in the
8.99\AA\ thick film. As the temperature increases the maximum effect moves
towards films of greater thickness before eventually disappearing. If one
correlates thickness with the sheet resistance of the films at the lowest
temperature, the maximum effect occurs near or above a resistance of 6300$%
\Omega ,$ which is very close to the quantum resistance for pairs. This is
very close to what one would judge to be the SI transition from examination
of Fig. 1. The actual SI transition may correspond to a film in the gap
between the 9.09\AA\ and 9.19\AA\ thick films.

We propose that the negative MR effect is associated with fluctuations in
the quantum critical region. Its magnitude would be expected to be a measure
of the strength of these fluctuations. The increase of the range of the
thicknesses over which the effect is seen, together with its weakening as
temperature is increased, is consistent with the boundaries of the quantum
critical region being determined by the condition $k_{B}T>\hbar \omega
_{c}\sim |d-d_{c}|^{\nu z}$, where $\hbar \omega _{c}$ is the energy scale
of the quantum fluctuations, $\nu $ is the correlation length exponent, and $%
z$ is the dynamical critical exponent \cite{Vojta}. Quantum critical
behavior would be expected to be cut off at high temperatures when $k_{B}T$
exceeds some microscopic energy scale in the problem. The negative MR effect
disappears at a temperatures above 0.3K. The thickness at which the maximum
effect is found shifts to thikcer films as T is increased. This shift would
imply that the crossover boundaries defining the quantum critical regime are
not symmetric.

Reports of negative magnetoresistance in disordered thin films and wires are
not new. Xiong, Herzog, and Dynes (XHD)\cite{Xiong} studied the behavior of
quench-condensed, homogeneous, amorphous thin film Pb wires. They reported a
low-field negative MR below the mean-field transition temperature with the
field transverse to the wire axis and perpendicular to the plane of the
film. Similar behavior was also reported by Markovi\'{c} and collaborators%
\cite{Markovic2} who studied MoGe wires grown on carbon nanotube substrates.
We focus the discussion on the work of XHD as details are available. XHD
suggested that the negative MR was enhanced by Coulomb correlations specific
to one-dimensional geometries. They further speculated that it might be the
result of negative superfluid density fluctuations close to the
superconductor-insulator (SI) transition. This was proposed by Kivelson and
Spivak\cite{KivelsonSpivak}. Apart from geometry being ultrathin films
rather than narrow ultrathin wires, there are a number of other differences
between the present work and that of XHD\cite{Xiong}. First the magnetic
field is {\it parallel} to the plane of the film whereas it is {\it %
perpendicular} to the plane in XHD and in the theory of Kivelson and Spivak%
\cite{KivelsonSpivak} . Second, the negative MR of XHD is found above 1K,
whereas in the present work, it exists only below 300mK. In XHD{\it \ }a
number of possible mechanisms for negative MR other than negative superfluid
density are considered and ruled out. Nonequilibrium charge imbalance
processes associated with phase slip centers can be excluded in the present
work as these processes are found in wires and not in films. Also the
effects we observe are found in the zero-current limit where the
current-voltage characteristic is linear. Phase slip centers would have
well-defined signatures in the current-voltage characteristics, which are
not seen. Another possibility raised by XHD relates to the quenching by the
magnetic field of spin fluctuations associated with electrons singly
occupying states in the {\it \ a}-Ge layer. (In their work it was actually
an {\it a-}Sb layer, but as shown by Hauser \cite{Hauser}, {\it a}-Sb and 
{\it a-}Ge are very similar in their properties. These localized electrons
were characterized in {\it a-}Ge using high-field calorimetry by van den
Berg and v. L\"{o}hneysen long ago \cite{Berg}.) With spin fluctuations
quenched by the magnetic field, superconducting fluctuations would be
enhanced, leading to a negative MR. In the present work, the range of fields
over which negative MR is growing extends to much higher values than those
at which spin fluctuations are suppressed using our previous argument. Thus
the negative MR observed is not likely due to the suppression of spin
fluctuations of localized electrons.

Another set of potentially relevant experiments are the studies of the SI
transition in perpendicular magnetic fields, where a peak, followed by
negative MR is found at fields larger than the critical field of the SI
transition. This was first observed in $In_{2}O_{3}$ films some years ago by
a Bell Laboratories group \cite{Hebard}, and has been reported more recently
by groups in Russia, Korea, Israel, and the US, respectively \cite
{Gantmakher,Lee,Baturina,Shahar,Kapitulnik2}. One might argue that the data
shown here is actually the same physics, but that the magnetic field scale
is dramatically reduced because the films are close to criticality with
regard to disorder. This is not likely to be the case. A feature of some of
the more detailed reports of a high-field resistance peak\cite
{Shahar,Kapitulnik2} is that the resistance in the region of the peak at
fixed magnetic field is described by $exp(T_{0}/T)$. This is not found in
our data. Also in our work the peak and the regime of negative MR disappear
above some film thickness and there is no trace of them in fields up to 12.5
T in films that are nominally superconducting.

There are a number of other models yielding negative MR in films, such as
the work of Beloborodov and collaborators \cite{Belob} and that of Galitski
and Larkin \cite{Galitski}. Since these involve perpendicular rather than
parallel magnetic fields they are not as relevant as the work of Lopatin,
Shah and Vinokur\cite{Lopatin}, although they may involve similar physics.
Yip \cite{Yip} proposed another mechanism for negative MR. He considered
superconductivity confined at a two-dimensional interface with strong
surface spin-orbit interaction and showed that an in-plane Zeeman field can
induce supercurrent flow. In other words, spin polarization induces
supercurrent flow. Although this calculation refers specifically to the
superconducting state, the idea might have relevance to the fluctuation
regime.

Although the calculations of Lopatin, Shah and Vinokur\cite{Lopatin} are not
specific to the present experimental geometry in that their quantum critical
point is approached by driving the transition temperature to zero with a
parallel magnetic field rather than by controlling disorder, the common
features of being close to the quantum critical point and the effect
occurring in parallel field, suggest that the underlying mechanism in that
calculation and the physics involved in the present work are likely to be
the same. A negative MR would then be a signature of critical fluctuations
associated with the quantum critical point and be an important diagnostic
for the SI transition. A detailed calculation relevant to the present
configuration would settle the issue.

The authors would like to thank N. Shah and A. Kaminev for useful
discussions. This work was supported by the National Science Foundation
Condensed Matter Physics Program under grant DMR-0138209.


\begin{figure}[tbp]
\caption[Evolution of$R(T)$ of the 9.19\AA\ film as a function of in-plane
magnetic field. Field values from top to bottom are: 12.5, 12, 11.6, 11.5,
11, 10, 9, 8, 7, 6, 5, 4, 3, 2, and 0 T. Inset: temperature at which dR/dT
becomes zero is plotted vs. applied field.]{Resistance as a function of
parallel magnetic field at 50mK (top curve), 100mK, 150mK, 200mK, 250mK, and
300mK (bottom curve) for the 9.05 \AA\ thick film. Data is shown at low
field to emphasize the negative magnetoresistance that appears at low
temperature. In fields higher than those shown, the $R(B)$ behavior is quite
linear.}
\label{fig2}
\end{figure}
\begin{figure}[tbp]
\caption{Magnetic field at the peak of $R(B)$curves vs. temperature for
8.99\AA , 9.05\AA , and 9.09\AA\ thick films from top to bottom. The line
corresponds to $\mu B/k_{B}T=1.$ }
\label{fig3}
\end{figure}
\begin{figure}[tbp]
\caption{Difference between resistances of the trough and peak of $R(B)$
curves $(R_{\min }-R_{\max })$as a function of thickness thickness d at
0.05K, 0.200K and 0.3K from bottom to top. }
\label{fig4}
\end{figure}


\begin{references}
\bibitem{Girvin}  S. L. Sondhi, S. M. Girvin, J. P. Carini, and D. Shahar,
Rev. Mod Phys. {\bf 69}, 315 (1997).

\bibitem{Mason}  N. Mason, and A. Kapitulnik, Phys. Rev. Lett.{\bf \ 82},
5341 (1999); N. Mason and A. Kapitulnik, Phys. Rev. B {\bf 65}, 220505
(2002).

\bibitem{Phillips}  Denis Dalidovich and Philip Phillips, Phys. Rev. Lett. 
{\bf 89}, 027001 (2002); Philip Phillips and Denis Dalidovich, Science {\bf %
302}, 243 (2003).

\bibitem{Trivedi}  Amit Ghosal, Mohit Randeria and Nandini Trivedi, Phys.
Rev. B{\bf \ 65}, 014501 (2001).

\bibitem{Lopatin}  A. V. Lopatin, N. Shah, and V. M. Vinokur,
cond-mat/0404623.

\bibitem{Hernand}  L. M. Hernandez and A. M. Goldman, Rev. Sci. Instrum. 
{\bf 73}, 162 (2002).

\bibitem{Strongin}  M. Strongin, R. S. Thompson, O. F. Kammerer,and J. E.
Crow, Phys. Rev. B {\bf 1}, 1078 (1970).

\bibitem{Chervenak}  J. A. Chervenak and J. M. Valles, Jr., Phys. Rev. B 
{\bf 61}, R9245, (2000).

\bibitem{Doniach}  D. Das and S. Doniach, Phys. Rev. B. {\bf 60}, 1261(1999).

\bibitem{Christiansen}  C. Christiansen, L. M. Hernandez, and A. M. Goldman,
Phys. Rev. Lett. {\bf 88}, 037004 (2002).

\bibitem{Markovic1}  N.Markovi\'{c}, A.M.Mack, G.Martinez-Arizala, C.
Christiansen, and A.M.Goldman, Phys. Rev. Lett. {\bf 81}, 701 (1998).

\bibitem{Matveev}  K. A. Matveev, L. I. Glazman, Penny Clarke, D. Ephron,
and M. R. Beasley, Phys. Rev. B {\bf 52}, 5289 (1995).

\bibitem{Mertes}  K. M. Mertes, D. Simonian, M. P. Sarachik, S. V.
Kravchenko, and T. M. Klapwijk{\it , }Phys. Rev. B {\bf 60}, R5093 (1999).

\bibitem{Vojta}  Matthias Vojta, Rep. Prog. Phys. {\bf 66}, 2069 (2003).

\bibitem{Xiong}  P. Xiong, A. V. Herzog, and R. C. Dynes, Phys. Rev. Lett. 
{\bf 78}, 927 (1997)

\bibitem{Markovic2}  N. Markovi\'{c}, C. N. Lau, and M. Tinkham, Physica C 
{\bf 387}, 44 (2003) and unpublished.

\bibitem{KivelsonSpivak}  B. Z. Spivak and S. A. Kivelson, Phys. Rev. B {\bf %
43}, 3740 (1991; S. A. Kivelson and B. Z. Spivak, Phys. Rev. B {\bf 45},
10490 (1992).

\bibitem{Hauser}  J. J. Hauser, Phys. Rev.B{\bf \ 11}, 738 (1975).

\bibitem{Berg}  R vandenBerg and H. v. L\"{o}hneysen, Phys. Rev. Lett. {\bf %
55}, 2463 (1985).

\bibitem{Hebard}  M. A. Paalanen, A. F. Hebard, and R. R. Ruel, Phys. Rev.
Lett. {\bf 69}, 1604 (1992).

\bibitem{Gantmakher}  V. F. Gantmakher, M. V. Golubkov, V. T. Dlgopolov, A.
A. Shashkin, and G. E. Tsydynzhapov, JETP Lett. {\bf 71}, 473 (2000); V. F.
Gantmakher, S. N. Eermolov, G. E. Tsydynzhapov, A. A. Zhukov, and T. I.
Baturina, JETP Lett. {\bf 77}, 424 (2003).

\bibitem{Lee}  Y. J. Lee, Y. S. Kim, E. N. Bang, H. Lim, and H. K. Shin, J.
Phys. Condens. Matter {\bf 13}, 8135 (2001).

\bibitem{Baturina}  Tatyana I. Baturina, D. R. Islamov, Z. D. Kvon, M. R.
Baklanov, and A. Satta, cond-mat/0210250.

\bibitem{Shahar}  G. Sambandamurthy, L. W. Engel, A. Johansson, and D.
Shahar, Phys. Rev. Lett. {\bf 92}, 107005 (2004).

\bibitem{Kapitulnik2}  Myles Steiner and Aharon Kapitulnik,
cond-mat/0406227; M. A. Steiner, G. Boebinger, and A. Kapitulnik,
cond-mat/0406232.

\bibitem{Belob}  I. S. Beloborodov and K. B. Efetov, Phys. Rev. Lett.{\bf \
82}, 3332 (1999); I. S. Beloborodov, K. B. Efetov, and A. I. Larkin, Phys.
Rev. B{\bf \ 61}, 9145 (2000).

\bibitem{Galitski}  V. M. Galitski and A. I. Larkin, Phys. Rev. B {\bf 63},
174506 (2001).

\bibitem{Yip}  S. K. Yip, Phys. Rev. B {\bf 65}, 144508 (2002).

\begin{figure}[tbp]
\caption{ Evolution of $R(T)$ for a series of eleven different thicknesses
of Bi. Film thicknesses are: 8.5 (top curve), 8.7, 8.8, 8.85, 8.91, 8.99,
9.05, 9.09, 9.19, 9.25, and 9.3\AA\ (bottom curve). }
\label{fig1}
\end{figure}
\end{references}
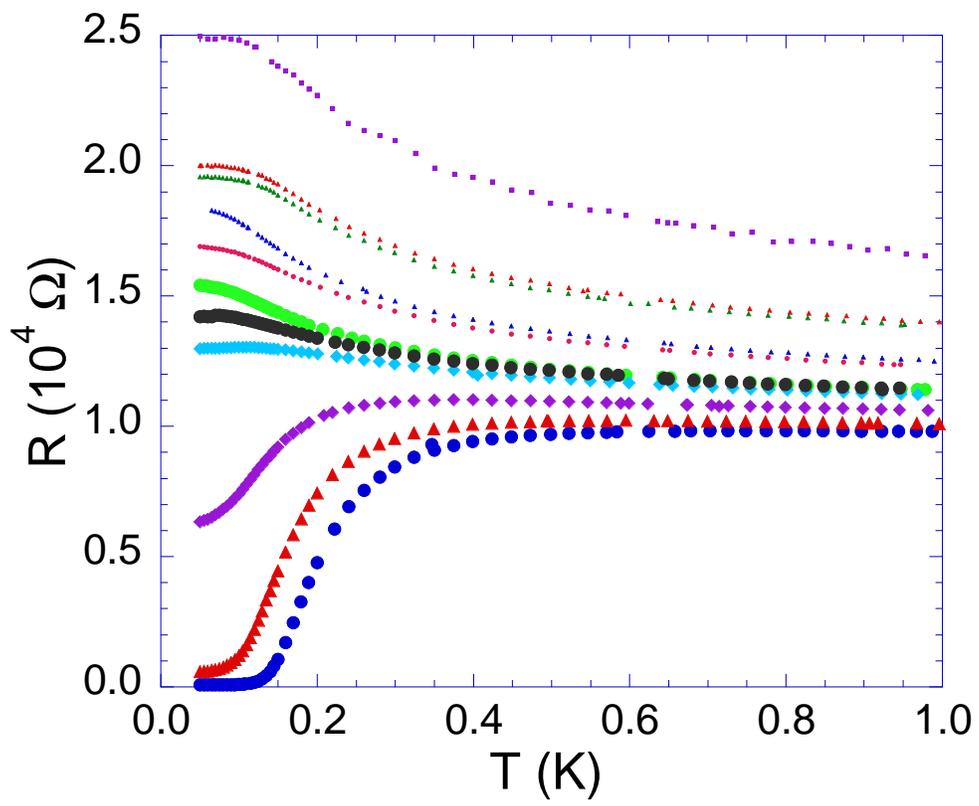
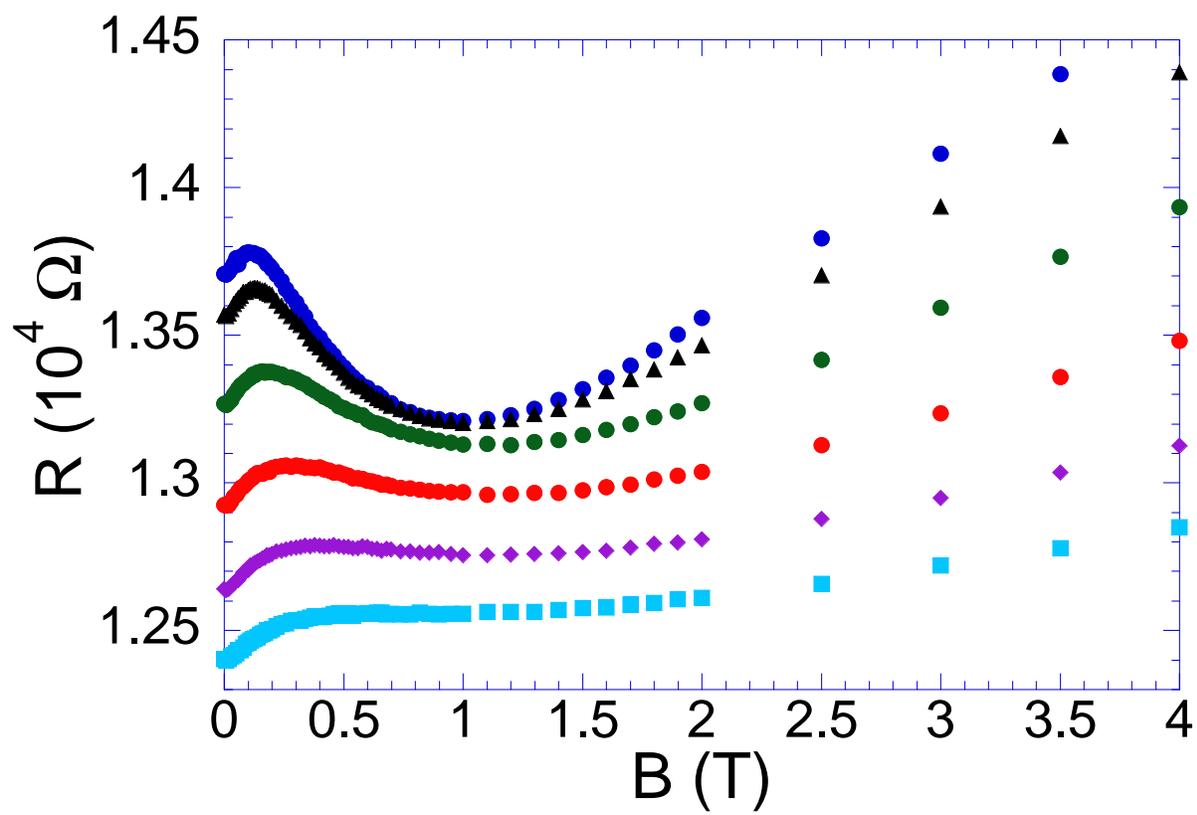
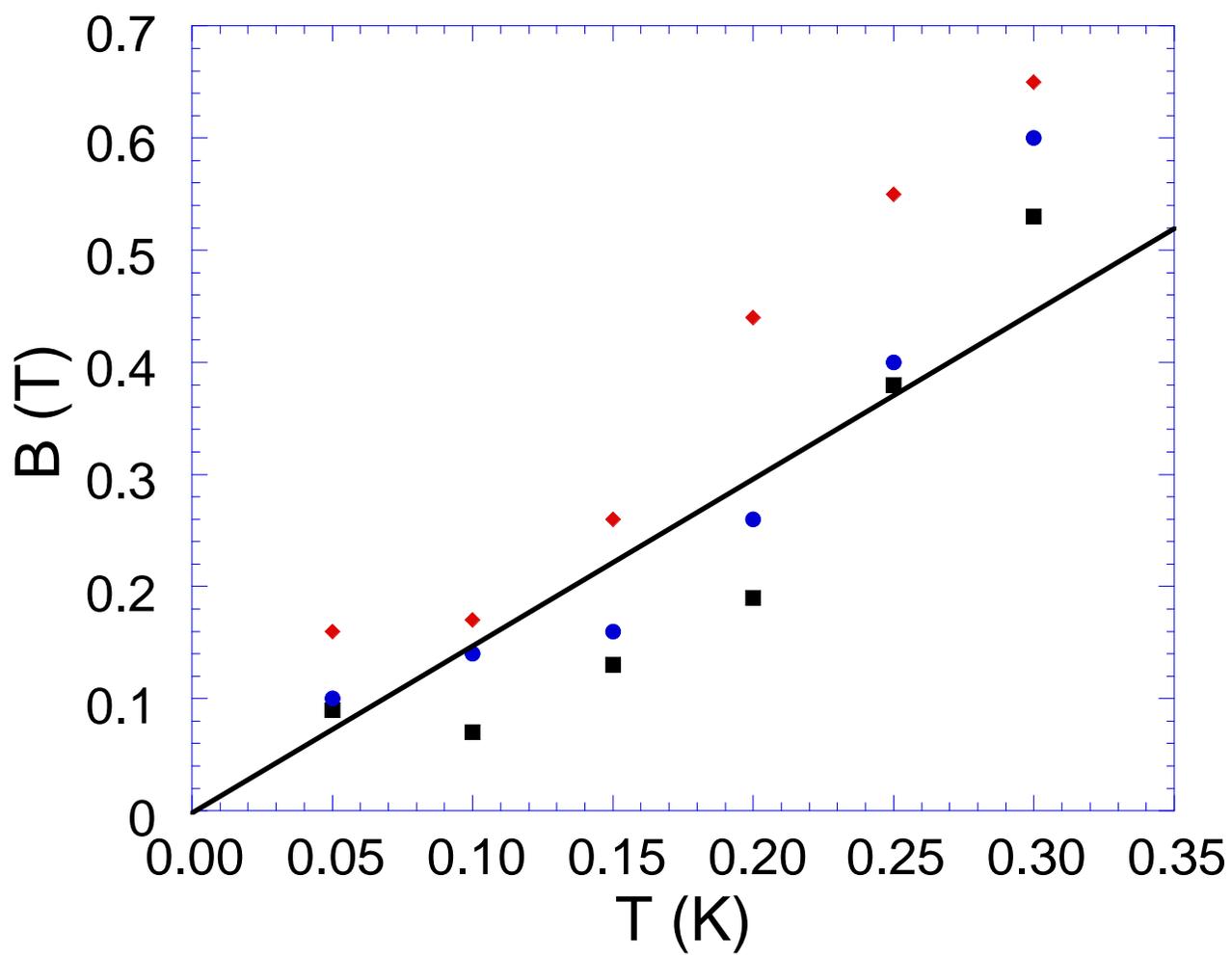
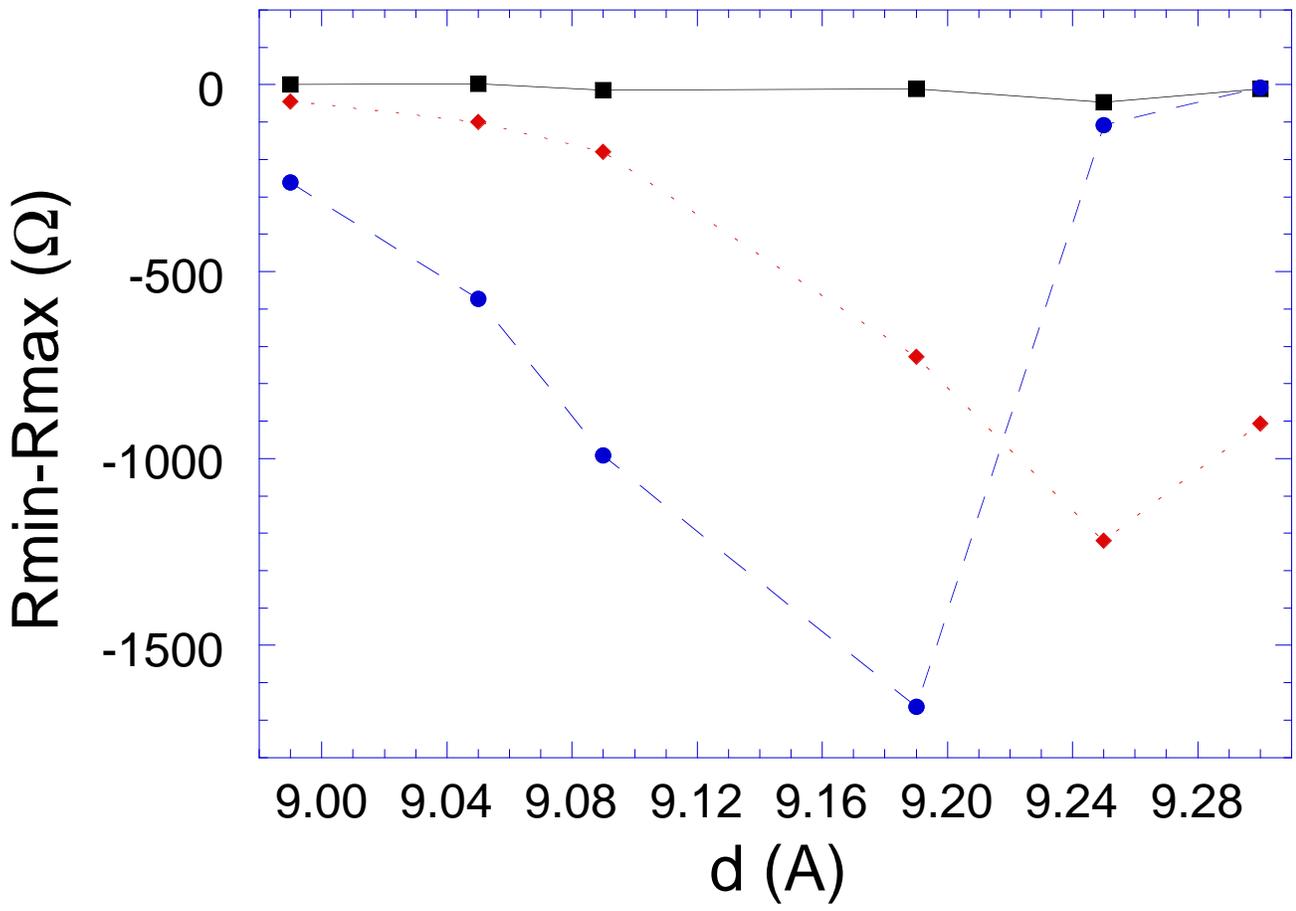
\end{document}